\documentclass[useAMS,usenatbib,usegraphicx,letters]{mn2e}
\usepackage{times}
\usepackage[fleqn]{amsmath}
\usepackage{amssymb}
\usepackage{amsbsy}

\newcommand{\zev}{\ensuremath{\mathcal{Z}}}

\title[21-cm foreground selection]{Selection between foreground models for global 21-cm experiments}

\author[G.\ J.\ A.\ Harker]{Geraint J.\ A.\ Harker\thanks{Email:
    g.harker@ucl.ac.uk}\\
    Department of Physics and Astronomy, University College London, London WC1E 6BT, UK
}

\begin{document}

\date{\today}

\maketitle

\begin{abstract}
The precise form of the foregrounds for sky-averaged measurements of the 21-cm line during and before the epoch of reionization is unknown. We suggest that the level of complexity in the foreground models used to fit global 21-cm data should be driven by the data, under a Bayesian model selection methodology. A first test of this approach is carried out by applying nested sampling to simplified models of global 21-cm data to compute the Bayesian evidence for the models. If the foregrounds are assumed to be polynomials of order $n$ in log--log space, we can infer the necessity to use $n=4$ rather than $n=3$ with $<2\ \mathrm{h}$ of integration with limited frequency coverage, for reasonable values of the $n=4$ coefficient.

Using a higher-order polynomial does not necessarily prevent a significant detection of the 21-cm signal. Even for $n=8$, we can obtain very strong evidence distinguishing a reasonable model for the signal from a null model with $128\ \mathrm{h}$ of integration. More subtle features of the signal may, however, be lost if the foregrounds are this complex. This is demonstrated using a simpler model for the signal that only includes absorption.

The results highlight some pitfalls in trying to quantify the significance of a detection from errors on the parameters of the signal alone.
\end{abstract}


\section{Introduction}\label{sec:intro}

The sky-averaged or `global' signal from the 21-cm line of hydrogen at redshifts $z\gtrsim 6$ has been put forward as a probe of reionization, the `cosmic dawn' (first stars and galaxies, $z\gtrsim 13$) and even the preceding `dark ages' at $z\gtrsim 30$ \citep{PRI12}. It may complement interferometric measurements of 21-cm fluctuations, and allow higher redshifts to be studied more quickly.

A persistent concern, however, is that it may not be possible to separate the 21-cm signal from bright foregrounds, which include diffuse synchrotron and free--free radiation from our Galaxy as well as emission from extragalactic sources \citep{SHA99}. The problem is more severe even than for interferometric measurements, since the features of the global 21-cm signal extend over many MHz, while fluctuations along individual sightlines in interferometric maps are expected to decorrelate over a bandwidth of $\lesssim 1\ \mathrm{MHz}$ \citep[e.g.\ ][]{BHA05,MEL06}. The global signal is thus likely to be more degenerate with the largely smooth foregrounds. Furthermore, it is difficult to obtain independent measurements of the foregrounds at high enough precision to be useful: interferometers may provide some insight but are sensitive to the spatially fluctuating part of the foregrounds \citep[though see][]{VED14pre,PRE15}, while monopole observations which are sensitive enough to detect the 21-cm signal are also likely to be the deepest and best calibrated foreground measurements at the appropriate frequencies. The foregrounds and 21-cm signal must therefore be inferred simultaneously from the data.

The degeneracy between the foregrounds and the signal may limit the usefulness of the fully blind component separation methods which have been applied to interferometric data \citep[e.g.\ ][]{CHA12,CHA13}. Instead, we might seek a framework which can incorporate stronger assumptions about the spatial structure and spectral smoothness of the foregrounds \citep{LIU13}. This raises the question of how restrictive our foreground models must be, or alternatively how complex a foreground model is required by the data.

In this letter, we adopt parametrized forms for the 21-cm signal and the frequency dependence of the foregrounds, and test whether the Bayesian evidence could be useful both for selecting an appropriate foreground model, and for inferring the presence of a 21-cm signal in the data given such a model. We consider synthetic data generated using only a highly simplified instrument model, but test whether preliminary measurements without the bandwidth or integration time of a full global signal experiment might be able to constrain the level of complexity present in the foregrounds. We then move on to consider constraints on the 21-cm signal itself, and the interplay between signal inferences and the order of the foregrounds.

The parameters of the signal, foregrounds and instrument are described in Sec.~\ref{sec:methods}. Here, we also briefly introduce the methods we use for computing the Bayesian evidence, and how the evidence is used to compare foreground models. The results of our evidence computations, for a range of different levels of foreground complexity and integration time, are given in Sec.~\ref{sec:res} and discussed further in Sec.~\ref{sec:disc}.

\section{Methods}\label{sec:methods}

\subsection{Signal and foreground modelling}\label{subsec:sigfg}

We consider an experiment which observes a single patch of sky for a length of time $t_\mathrm{obs}$. The noise on the measurement is purely thermal noise computed according to the radiometer equation, assuming an antenna with a flat frequency response and an efficiency of 85 per cent. In addition to the sky noise, there is a contribution from the receiver which we take to be $226.2\ \mathrm{K}$, though this is not intended to be representative of any particular experiment. We also assume that the only foregrounds present are smooth, diffuse Galactic foregrounds, and a sea of extragalactic sources that appear as a diffuse foreground at the resolution of proposed 21-cm global signal experiments. That is, we neglect sources such as the Sun and Moon, and contamination by anthropogenic radio frequency interference. This is motivated partly by the possibility of missions such as the {\it Dark Ages Radio Explorer} \citep[\textit{DARE};][]{BUR12}, which would avoid most foregrounds by taking data only while in a low orbit over the far side of the Moon, and partly by a desire to reduce the number of parameters in the model, in order to make large numbers of evidence computations feasible.

The diffuse foregrounds are taken to have the form of a polynomial in $\ln(T)$--$\ln(\nu)$, i.e.
\begin{equation}
  \ln T_\mathrm{FG}=\ln T_0 + \sum_{i=1}^na_i[\ln(\nu/\nu_0)]^i\ ,
  \label{eqn:fgform}
\end{equation}
where $\nu_0=80\ \mathrm{MHz}$ is an arbitrary reference frequency, and $\{T_0,a_1,a_2,\ldots,a_n\}$ are the parameters of the model. By increasing $n$, we can study progressively more complex, less smooth foregrounds.

Where we include the 21-cm signal, it is parametrized as a cubic spline passing through a number of maxima and minima (turning points) following \citet{PRI10}. The frequency and brightness temperature of these turning points are the parameters of the signal model. We restrict our attention to frequencies of $35$--$120\ \mathrm{MHz}$, and so we only fit the parameters of \citeauthor{PRI10}'s turning points 1--3 (corresponding to the start of Ly$\alpha$ pumping, the start of effective heating, and signal saturation, respectively), leaving the position of turning point 0 (in the true dark ages) and 4 (the end of reionization) fixed. We refer below to turning points 0--4 as A--E, respectively, to avoid confusion with subscripts.

When we simulate the noisy spectrum, we assume that $\{T_0,a_1,a_2,a_3\}=\{2039.611,-2.42096,-0.08062,0.02898\}$, computed by fitting a quiet region of the global sky model (GSM) of \citet{dOC08}, convolved with a beam with a full width at half-maximum of $72^\circ$, with a third-order polynomial over $35$--$120\ \mathrm{MHz}$. Higher order coefficients are varied as described in Sections \ref{subsec:fginf} and \ref{subsec:siginf}. For the signal, we assume the same turning point positions as the fiducial model of \citet{PRI10}. This leads to the input signal shown in Fig.~\ref{fig:c_pp}.

\subsection{Evidence computation}\label{subsec:evcomp}

We use a slightly modified version of \textsc{multinest} v3.2 (\citealt{FER08}; \citealt*{FER09}; \citealt{FER13}), which implements nested sampling \citep{SKI04} to compute the Bayesian evidence, \zev. This also yields weighted samples of the posterior probability distribution of the parameters given the data. Uniform priors are used for the turning point frequencies and positions. For $T_0$, we assume a Gaussian prior with mean and standard deviation equal to the `true' $T_0$, but truncated at zero. For $a_1$ (the spectral index at $\nu_0$), we assume a Gaussian prior with a mean of the `true' $a_1$ and a standard deviation of 0.1, while for all other $a_i$ we use a Gaussian prior with mean 0 and standard deviation of 0.1. We adopt this value because fourth-order polynomial fits to the GSM yield values of $a_4$ between $-0.024$ and $0.037$ in individual pixels, but this reduces by around an order of magnitude after smoothing with a beam of a typical size for global 21-cm experiments. We would expect higher order terms to be smaller.

\section{Results}\label{sec:res}

\subsection{Foreground inference with limited frequency coverage}\label{subsec:fginf}

\begin{figure}
  \begin{center}
    \leavevmode
    \includegraphics[width=8cm,clip=true]{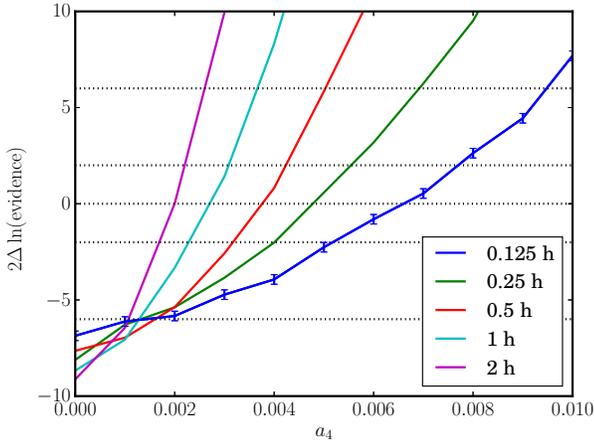}
    \caption{We show $2[\ln(\zev_4)-\ln(\zev_3)]$, where $\zev_3$ and $\zev_4$ are the evidence for a third and fourth-order polynomial foreground model, respectively, for the simulation setup described in Sec.~\ref{subsec:fginf}. This is plotted as a function of $a_4$, the coefficient of the fourth-order term in the foreground model. The different curves are for observations with different amounts of observing time, as shown. Error bars are shown only on the $t_\mathrm{obs}=0.125\ \mathrm{h}$ curve, for clarity, but the errors for the other curves are similar. Dotted lines at $2\Delta\ln\zev=0,2,6$ (and the axis limits at 10) show typical thresholds used to assess the degree to which the data favour one model over another \citep{KAS95}.}\label{fig:ev_r}
  \end{center}
\end{figure}

We start with a test which considers only foregrounds. Data are simulated using the instrument model described in Sec.~\ref{subsec:sigfg}, but only in the ranges 40--50, 75--85 and 110--120 MHz, though we do fit all three segments of the spectrum simultaneously. For this test, we consider only short integration times. This emulates an experiment with limited scope and which does not attempt to cover the whole range from 40--120 MHz with a single antenna having a smooth frequency response, which is technically challenging. $\{T_0,a_1,a_2,a_3\}$ remain fixed, but we simulate data for different values of $a_4$ (between zero and 0.01 in steps of 0.001). In each case, we attempt to fit the data using a third order and a fourth order polynomial model, in order to test whether or not the addition of $a_4$ to the parameter set is justified by the Bayesian evidence. The results, expressed in terms of the evidence ratio (or difference in log-evidence, $\Delta\ln\zev$) between the third and fourth order models, are shown in Fig.~\ref{fig:ev_r}. Differences in $2\Delta\ln\zev$ of 2,6 and 10 correspond to borders between the categories of `not worth more than a bare mention', `positive', `strong' and `very strong' evidence for one model over the other, according to the guidelines of \citet{KAS95}.

If $a_4=0.01$ (note that $|a_4|$ is larger than this for many pixels in our GSM), we achieve strong evidence for a non-zero $a_4$ in only 7.5 min of integration. With $t_\mathrm{obs}=1\ \mathrm{h}$, we obtain very strong evidence against the simpler, third-order model for $a_4>0.004$. That is, reasonable levels of foreground complexity can be constrained with a brief observation of limited frequency coverage (assuming it is sufficiently well calibrated), much less than is required to detect the 21-cm signal.

Note also that for $a_4=0$, the evidence always (correctly) favours the third-order foreground model, with longer integrations producing stronger evidence for the simpler model. The evidence never becomes conclusive, however, even for the very long integrations (not shown) we have run as test cases. It seems unlikely that we could confidently use only a third-order model for the foregrounds in a full 21-cm experiment.

\subsection{Signal inferences with complete frequency coverage}\label{subsec:siginf}

\begin{figure}
  \begin{center}
    \leavevmode
    \includegraphics[width=8cm,clip=true]{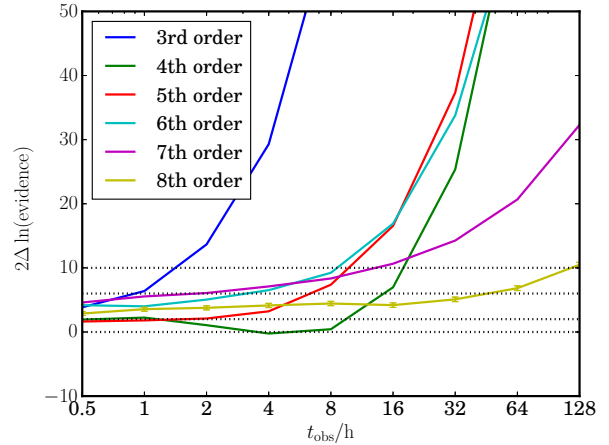}
    \caption{The evidence for the presence of a signal in the data, $2[\ln(\zev_\mathrm{with})-\ln(\zev_\mathrm{without})]$, for different levels of foreground complexity, is shown as a function of the integration time, $t_\mathrm{obs}$. In computing $\zev_\mathrm{with}$ we assume a 21-cm signal, parametrized by its turning points, is present in the data, while in computing $\zev_\mathrm{without}$, no signal is included in the model. In all cases, the `correct' order, $n$, of the foreground model was assumed. In generating the synthetic data, we used $a_n=0.001$, and $a_i=0$ for $3<i<n$, and the experimental setup described in Sec.~\ref{subsec:siginf}. Error bars are shown only on the eighth-order model curve, for clarity, but are similar for all the other curves. Dotted lines show typical thresholds in $2\Delta\ln\zev$ used to assess the degree to which the data favour one model over another.}\label{fig:ev_r_sig}
  \end{center}
\end{figure}

We now move on to considering a more ambitious but challenging experiment aimed at detecting the 21-cm signal itself. We simulate data over the complete range from $35$ to $120\ \mathrm{MHz}$ and include the cosmological signal in the simulations. In Fig.~\ref{fig:ev_r_sig}, we test how much integration time is required to obtain a detection in the presence of different levels of foreground complexity. $\{T_0,a_1,a_2,a_3\}$ remain fixed, as before, but the order $n$ of the simulated foregrounds is varied. We take $a_n=0.001$, and $a_i=0$ for $3<i<n$, and the fitting is done using the same $n$ as the simulation. We plot $2\Delta\ln\zev$ between a model including a 21-cm signal and one without, as a function of integration time, for $3\leq n\leq 8$.

With only third-order foregrounds, there is very strong evidence for the presence of a 21-cm signal within $2\ \mathrm{h}$ of integration, but such a low-order model is likely to be unrealistic \citep{VED14}. All higher orders require at least $8\ \mathrm{h}$. The increase does not appear to be monotonic, with $n=4$ requiring more time than $n=5$ or $6$. This anomaly comes about because the foreground with $a_4=0.001$ partially mimics the signal, which thus requires more time to distinguish. We have confirmed this by rerunning the $n=5$ and $n=6$ simulations using $a_4=0.001$ rather than $a_4=0$, in which case the increase becomes monotonic, as expected. This highlights the fact that even if the parametrized model for the foregrounds is correct, it is possible to be unlucky with the values these parameters take, increasing the time required for a detection.

For $n>6$, the behaviour changes, and the increase in $\Delta\ln\zev$ with $t_\mathrm{obs}$ becomes less steep, perhaps suggesting that degeneracies between the foreground and signal models are starting to become more important. None the less, with $t_\mathrm{obs}=128\ \mathrm{h}$, there is significant evidence for a signal even with $n=8$, the order which \citet*{BER15} found was required to extract unbiased estimates of the signal parameters in their modelling, which included structure introduced by the antenna response.

\begin{figure}
  \begin{center}
    \leavevmode
    \includegraphics[width=8cm,clip=true]{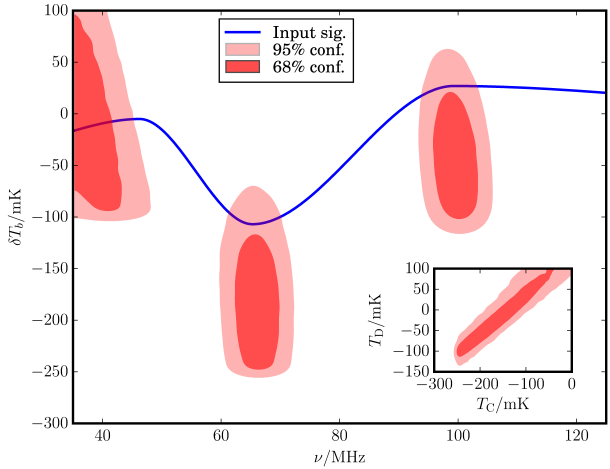}
    \caption{Constraints on the positions of the three fitted turning points (maxima and minima) in the 21-cm signal model for a third-order polynomial foreground model and $128\ \mathrm{h}$ of integration, with the experimental setup described in Sec.~\ref{subsec:siginf}. The blue curve shows the input 21-cm signal spectrum used to generate the synthetic data, while the red, filled contours show the 68 and 95 per cent credible regions on the frequency and amplitude of each turning point. The inset shows the very strong correlation between the inferred amplitude of the two higher frequency turning points (C and D), which shows that the signal is quite well constrained up to an overall additive normalization. Although the evidence for the presence of a signal is overwhelming (2$\Delta\ln\zev = 1312$; recall that values $>10$ are considered `very strong' evidence), the amplitude of turning point C (the deepest point of the trough in the middle of the band) is only a few standard deviations away from zero, demonstrating the fact that this is a poor measure of the significance with which a signal is detected. The odd shape of the contours at low $\delta T_\mathrm{b}$ is because they are cut off by the priors: lower values would be unphysical since they would imply a Universe cooling faster than adiabatically.}\label{fig:c_pp}
  \end{center}
\end{figure}

Even a very significant detection may yield parameter constraints which are not straightforward to interpret, however, as we show in Fig.~\ref{fig:c_pp}. Here, we show the credible regions for the position (in frequency and brightness temperature) of the three turning points lying in our frequency range, overlaid on a plot of the input signal. The constraints are taken from the $t_\mathrm{obs}=128\ \mathrm{h}$ realization with $n=3$, so the 21-cm signal is detected conclusively.

The frequency of each turning point is measured reasonably well, apart perhaps from the low-frequency turning point at $46.2\ \mathrm{MHz}$, for which the contours do not close within our frequency range, suggesting that only an upper limit could be measured. The amplitude errors are large, however. For example, the brightness temperature of the absorption minimum, the largest and most easily detected feature in the signal, is measured at $-176.6\pm 42.3\ \mathrm{mK}$. This might be labelled a `4--$\sigma$ detection', even though the Bayesian evidence suggests that a 21-cm signal is present at much greater confidence. One reason is the difficulty in constraining the overall zero-point of the signal, as demonstrated by the inset panel in Fig.~\ref{fig:c_pp}. This shows the joint constraints on the brightness temperature of this feature and that of the emission maximum corresponding to the start of reionization. The difference between these two temperatures (and thus the overall shape of the signal) is measured much more precisely than either on its own. Care is therefore required in translating measurements of the parameters of a fitting function to physical quantities of interest, as in e.g.\ \citet*{MIR13}; ideally, we would constrain the parameters of a physical model directly, rather than passing through an intermediate fitting function.

\begin{figure}
  \begin{center}
    \leavevmode
    \includegraphics[width=8cm,clip=true]{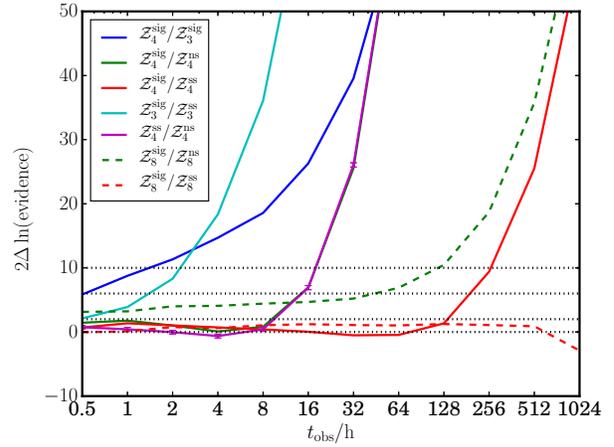}
    \caption{The interplay between inference of the foreground order and the signal model. For all the solid lines, data are simulated for a fourth-order polynomial model with $a_4=0.001$, and the full signal model. Different lines show $2\ln r$ where $r$ is the evidence ratio given in the legend. The subscripts of \zev\ in the legend show the polynomial order used to fit the foregrounds, while the superscripts label three different signal models: a null signal model (ns), a simple signal model (ss) in which the amplitude of turning point D is fixed to be zero, and the full signal model (sig). Note that the $\zev^\mathrm{ss}_4/\zev^\mathrm{ns}_4$ line with error bars almost overplots the $\zev^\mathrm{sig}_4/\zev^\mathrm{ns}_4$ line. The two dashed lines are analogous to the two solid lines of the same colour: in each, the data are simulated for eighth-order foregrounds with the full signal, and we show $2\ln r$ for the cases given in the legend.}\label{fig:ev_r_3_4_7_8}
  \end{center}
\end{figure}

In Fig.~\ref{fig:ev_r_3_4_7_8}, we show the effect of integration time on simultaneous constraints of the foreground order and the signal model. As part of this, we introduce a simplified signal model in which the signal never goes into emission: the high-frequency maximum in the signal (turning point D, at $\sim 100\ \mathrm{MHz}$ in Fig.~\ref{fig:c_pp}) has its amplitude fixed to zero, so that the signal is zero at all higher frequencies. The frequency at which the signal reaches zero is still allowed to vary. In this model, the intergalactic medium reionizes while it is still cold. Since only an absorption trough is present, it is somewhat similar in spirit to the Gaussian signal model used by \citet{BER15}. For the solid lines in Fig.~\ref{fig:ev_r_3_4_7_8}, the data are simulated with fourth-order foregrounds, while for the dashed lines they are simulated with eighth-order foregrounds. The subscripts in the legend show the polynomial order assumed in the fit.

The blue, solid line shows the evidence ratio between fits assuming fourth-order and third-order foregrounds, when the data are simulated with the fiducial 21-cm signal and with $a_4=0.001$, and where the parameters of the full signal are fitted for along with the foreground parameters. The inclusion of the signal does not prevent a significant detection of foreground complexity, for which there is very strong evidence with $<2\ \mathrm{h}$ of integration. If no signal is present or assumed, the evidence is very strong even for $0.5\ \mathrm{h}$, so we do not include these lines in order to avoid having to compress the scale of the plot.

The solid green line is identical to the line of the same style in Fig.~\ref{fig:ev_r_sig}, and shows how well the presence of a signal can be inferred for fourth-order foregrounds. It is almost overplotted by the magenta line, which shows the evidence ratio between a fit including the simple signal (ss) model and the null model (note that the data were still simulated assuming the full signal model). This similarity shows that assuming a slightly incorrect signal model may not be too harmful to a detection. The dashed green line reproduces the yellow line from Fig.~\ref{fig:ev_r_sig} and shows again the effect of increasing the order of the foreground model on the integration time required for a detection.

Distinguishing the full signal from a simple signal is much more difficult, however. The cyan, solid red and dashed red lines show the evidence ratio between a fit using the full signal (including the emission maximum) and one using the simple signal (absorption only), for $n=3$, $4$ and $8$, respectively. For $n=3$, the full signal is very strongly favoured over the simple signal within $4\ \mathrm{h}$. This detection may be spurious however, since for $n=4$ (the `correct' order) it requires more than $256\ \mathrm{h}$ to achieve. Care is clearly required in choosing an appropriate foreground model. For $n=8$, meanwhile, even $1024\ \mathrm{h}$ are insufficient to distinguish the full signal from the simple signal. The absorption trough is clearly the outstanding feature in our fiducial model. To detect more subtle features in the signal, such as a broad emission maximum expected in many reionization models, long integrations will be necessary, but not sufficient. Chromatic effects from the instrument, which led to the eighth-order fits required by \citet{BER15}, or effects from the ionosphere, will also have to be very tightly controlled or eliminated.

\section{Discussion and conclusions}\label{sec:disc}

In this letter, we have started to make the case for applying a Bayesian model selection methodology to the measurement of the foregrounds and the cosmological signal in global 21-cm experiments. This will allow the data to inform us about the appropriate level of complexity for our foreground model, and provides a more rigorous means of quantifying our confidence in any detection of the 21-cm signal. We have applied nested sampling to simplified realizations of spectra from global 21-cm measurements, using a polynomial foreground model (in log--log space) and a simple parametric form for the 21-cm signal. The framework is easily able to incorporate other components and different parametrizations, however. For example, it can be used to select between 21-cm signal models, rather than simply distinguishing them from a null signal.

If this methodology is applied to observational data, it may be necessary to include instrumental parameters, and to deal with spectra from multiple sky regions simultaneously, as in the Markov Chain Monte Carlo approach of \citet{dare_mc}. This will greatly increase the number of parameters required to describe the data. This increase in dimensionality is especially concerning given the exponential scaling of computational cost with number of parameters found by \citet{ALL14}, and has caused us problems in extending our analysis to multiple sky regions using \textsc{multinest}. Different algorithms to compute the evidence may be required. For ground-based 21-cm experiments, we may also need to include terms for the emission and absorption from the ionosphere in our model \citep[e.g.\ ][]{ROG14}.

Constraining the parameters of a physical model directly, rather than using a simple fitting function for the 21-cm signal, also increases the computational requirements (despite the development of efficient codes, e.g.\ \citealt{MIR14}), and raises the question of whether there might be a better parametrization than the `turning points' model used here, in the sense of being less degenerate with the foreground model while retaining the maximum amount of information about physical quantities. This is a topic of ongoing study.

\section*{Acknowledgements}

I acknowledge funding from: the People Programme (Marie Curie Actions) of the European Union's Seventh Framework Programme (FP7/2007--2013) under REA grant agreement no.\ 327999; the LUNAR consortium, which was funded by the NASA Lunar Science Institute (via Cooperative Agreement NNA09DB30A) to investigate concepts for astrophysical observatories on the Moon; and additional support provided by the Director's Office at the NASA Ames Research Center. This work utilized the Janus supercomputer, which is supported by the National Science Foundation (award number CNS-0821794) and the University of Colorado Boulder. The Janus supercomputer is a joint effort of the University of Colorado Boulder, the University of Colorado Denver and the National Center for Atmospheric Research. I would like to thank Emma Chapman and Jack Burns for stimulating discussions related to this letter, and the referee for suggesting the tests which led to Fig.~\ref{fig:ev_r_3_4_7_8} and for helping to improve the clarity of the paper elsewhere.

\bibliography{ns_fg_bib}

\end{document}